\documentclass{article}
\usepackage{algorithm}
\usepackage{algpseudocode}
\usepackage{amsmath}
\usepackage{amssymb}
\usepackage{times}

\newcommand{\Offset}{D}
\newcommand{\PriorCov}{\Sigma(\sigma)}
\newcommand{\ModelPred}{\Pi}
\newcommand{\Predict}{\texttt{Predict}}
\newcommand{\ScaledPredict}{\texttt{ScaledPredict}}
\newcommand{\SumBy}{\texttt{SumBy}}
\newcommand{\GaussPredict}{\texttt{GaussPredict}}
\newcommand{\PriorMarginal}{\texttt{PriorMarginal}}
\newcommand{\GaussPriorMarginal}{\texttt{GaussPriorMarginal}}

\newcommand{\dgamma}{\texttt{dgamma}}
\newcommand{\dnorm}{\texttt{dnorm}}
\newcommand{\RgdB}{B}

\newcommand{\BExp}[2]{B^{#2}_{#1}[I]}

\newcommand{\invvar}{\text{invvar}}
\newcommand{\error}{\text{error}}
\newcommand{\events}{\text{events}}
\newcommand{\pevents}{\text{pevents}}

\newcommand{\new}{\text{new}}

\newcommand{\argmax}[1]{\underset{#1}{\operatorname{argmax}}}

\begin{document}

\title{A Scalable Blocked Gibbs Sampling Algorithm For Gaussian And Poisson Regression Models}
\author{Nicholas A. Johnson\footnote{e-mail:naj@google.com}, Frank O. Kuehnel, Ali Nasiri Amini\\
		Google Inc}
\date{\today}
\maketitle

\begin{abstract}
Markov Chain Monte Carlo (MCMC) methods are a popular technique in Bayesian
statistical modeling.  They have long been used to obtain samples from posterior
distributions, but recent research has focused on the scalability of these techniques
for large problems. We do not develop new sampling methods but instead describe a blocked Gibbs sampler
which is sufficiently scalable to accomodate many interesting problems.  The sampler
we describe applies to a restricted subset of the Generalized Linear Mixed-effects Models 
(GLMM's); this subset includes Poisson and Gaussian regression models.  The 
blocked Gibbs sampling steps jointly update a prior variance parameter 
along with all of the random effects underneath it. We also discuss extensions
such as flexible prior distributions.\end{abstract}

\section{Introduction \label{sec:intro}}
There has been a great deal of work on implementing efficient, large-scale
regularized regressions, but there has been much less progress in scaling up
fully Bayesian regression models.  Bayesian and Empirical Bayes models (such as
GLMM's) have found wide use when applied to smaller datasets.  Two of
the more popular software packages are STAN \cite{stan2015rstansoftware} and
the lme4 R package \cite{bates2014fitting}.  STAN is highly-customizable and
it uses MCMC to draw posterior samples of model parameters.  The lme4 software is perhaps the most popular R
package for mixed effects models, but its implementation is based on Laplace
approximation which involves factorization of large matrices.

Recent work to scale up Bayesian model inference includes consensus Bayes
\cite{scott2013bayes}, stochastic gradient Langevin dynamics (SGLD)
\cite{welling2011langevin}, and the Weierstrass sampler \cite{wang2013parallelizing}.  
The intent of our publication is to present blocked Gibbs sampling in terms of
simple, scalable operations for solving large regression models.  The narrow
class of Bayesian regression models that we consider also have associated MCMC
moves which depend on a (relatively) small set of sufficient statistics.

In this section we first discuss an example of the type of data that our algorithm 
can model.  We then give some background on random effects models  and follow by describing the class of GLMM's
that our algorithm applies to.  Finally, we end this section with a review of 
background material and related work.

The data we accept as input can be summarized through a matrix such as that
in table (\ref{tab:exampledata1}).  When modeling that example data, our goal
could be to predict say the number of actions users took, \texttt{n.actions}, 
based on features \texttt{url} and \texttt{ad.id} and using \texttt{n.views} as an offset (the
number of ads the user saw).  In practice we would have many more feature
columns in our dataset.  These features often have a ``long tail" of levels
with little associated data, and Bayesian priors or regularization are essential
in order to make use of them.  Count data is common in applications, and we focus
on it for most of this report; however, section \ref{sec:GaussReg} briefly
discusses Gaussian models as well.

\begin{table}[position specifier]
  \centering
  \begin{tabular}{ | l | c | c | r | }
    \hline
    n.views & n.actions & url & ad.id \\ \hline \hline
    52 & 4 & abc.com & 83473 \\ \hline
    73 & 5 & xyz.edu & 40983 \\ \hline
    19 & 0 & abc.com & 4658 \\ \hline
    532 & 16 & efg.com & 40983 \\ \hline
    3 & 0 & z.com & 4658 \\ \hline
    ... & ... & ... & ... \\ \hline
    \hline
  \end{tabular}
  \caption{An example dataset for which n.actions could be modeled as a Poisson
count depending on two features, ``url" and ``ad.id", and on an observed offset ``n.views".}
  \label{tab:exampledata1}
\end{table}

We will start by describing GLMM's as they are usually presented and then discuss
how the data in table (\ref{tab:exampledata1}) could be summarized in terms of
this notation. A Poisson GLMM can be written as:
\begin{equation}
\{Y|b,\beta, \Offset\} \sim \text{Pois}(\Offset\exp(X\beta + Zb) )
\end{equation}
where $Y \in \mathbb{N}^n$ is the Poisson response, $X \in \mathbb{R}^{n\times p}$
is the fixed effects model matrix, $\beta \in \mathbb{R}^p$ is the fixed effect parameter,
$Z \in \mathbb{R}^{n\times r}$ is the random effect model matrix,
$b \in \mathbb{R}^r$ is the random effect (a random variable), and
$\Offset \in \mathbb{R}_{>0}^n$ is a positive offset.  Typically a
Gaussian prior is placed on $b$: $b \sim N(0, \PriorCov)$ i.e. $b$ has a normal
distribution with covariance $\Sigma$ and $\Sigma$ is in
turn parameterized by a lower-dimensional vector $\sigma$.

A common special case is that $\sigma \in \mathbb{R}_{\geq 0}^F$ where $F$ is the number of families (i.e.
feature columns in the initial example), and $\PriorCov$ is a
diagonal matrix with $\sigma_1$ appearing on the first $L_1$ diagonal elements,
$\sigma_2$ on the next $L_2$, and so on (and $L_1+...+L_F = r$).  In the language
of the ``lme4" R package \cite{bates2014fitting} a Poisson model of the data in table (\ref{tab:exampledata1})
can be specified as:
\begin{equation}
 \texttt{n.actions} \sim 1 + (1|\texttt{url}) + (1|\texttt{ad.id}) + \texttt{offset}(\log(\texttt{n.views}))
\end{equation}

When modeling this example dataset we would have $F=2$ random effect families,
$p = 1$ fixed effects (so $X$ is the $n\times 1$ matrix $[1]_{n\times 1}$),
$L_1$ would be the number of unique url's in the dataset,
and $L_2$ would be the number of unique \texttt{ad.id}'s.  Then $\sigma_1^2$
is the prior variance of the url random effects and $\sigma_2^2$ is the prior
variance of the \texttt{ad.id} random effects.

To be more precise, we define the index set $J_k$ to be the
$k$'th block of indices: $\{T_{k-1} + 1, ..., T_{k-1} + L_k \}$ where
$T_k := \sum_{j=1}^{k}L_j$ and $T_0=0$.  The random
effect variance matrix has $(j,j)$'th diagonal element $\PriorCov_{jj} = \sigma_{k}^2$ when $j \in J_k$.

The mixed effects model is then fit by maximizing the marginal likelihood:
\begin{align}
  \ell(\beta, \sigma) &:= \int P(\{b_i\}_{i=1}^r|\sigma)\left[\prod_{j=1}^n P(Y_j|\Offset_j, X_j, Z_j, \beta, b) \right] db_1...db_r \label{eqn:mixefflik} \\
\text{where } P(\{b_i\}_{i=1}^r|\sigma) &= \prod_{f=1}^F \prod_{j=1}^{L_f}P(b_{T_{f-1}+j}|\sigma_f)
\end{align}
where $X_j$ and $Z_j$ denote the j'th rows of the respective matrices,
and $\Offset_j$ and $Y_j$ refer to the j'th element of the respective vectors.

Usually the integral in (\ref{eqn:mixefflik}) cannot be computed in closed form, so it is approximated by MCMC
or Laplace approximation \cite{bates2014fitting}.  Gaussian models are an exception
but even in that case calculating the marginal likelihood can involve matrix factorizations
which are prohibitive to compute.  Software such as lme4 can handle
more general prior covariance structure than that described above, but we focus
on the case where $\Sigma$ is diagonal.

When modeling Poisson data, the algorithm will only apply to the restricted case
that: (a) $p=1$, $X = [1]_{n\times 1}$, (b) $Z$ is a 0-1 matrix (i.e. $Z_{jk} \in \{0, 1\}$),
 and (c) the partial row sums within a family's block of columns are equal to one: $\sum_{t \in J_k}Z_{it} = 1$.
It is easy to instead accommodate $\sum_{t \in J_k}Z_{it} \in \{0, 1\}$, but we
omit the details.  When modeling Gaussian data we relax this to allow any real valued
entries but maintain a restriction on the sparsity pattern:
$\sum_{t \in J_k}1\{Z_{it} \neq 0\} = 1$ for $k=1,2,...,F$

Condition (c) states that we have conditional independence between the random
effects parameters $b$ within a single feature family ``k''.  We also depend on this sparsity
pattern to store $Z$ efficiently.  We discuss this in more detail in section
\ref{sec:GammaPoisson}.  Next we choose a Gamma prior distribution instead
of the standard log-normal one for GLMM’s:
$\exp(b_j) \sim \text{Gamma}(\sigma_{k}^{-2}, \sigma_{k}^{-2})$, where
$E[\exp(b_j)] = 1$ and $Var(\exp(b_j)) = \sigma_{k}^{2}$ for $j \in J_k$.
The mean-one restriction is for identifiability; without it the Gamma prior
distributions could be scaled by any amount and an adjustment to the fixed
effect would yield exactly the same data distribution. We will see that
the conjugacy of the Gamma and Poisson distributions simplifies some sampling
steps in our algorithm.

With these restrictions and conjugate priors in place, we develop a blocked Gibbs sampling 
iteration in which we update a single family at a time.  We jointly update a
prior parameter and all of the random effects beneath it.

We end this section with a review of background material and related work.  Gibbs
sampling is widely used, so we just cover some early papers, relevant textbooks,
and related applications to Bayesian regression models.

Gelfand and Smith \cite{gelfand1990sampling} introduced the statistics community
to Gibbs sampling as a computational technique for inference in Bayesian models.
This paper did not propose Gibbs sampling (they attributed it to \cite{geman1984stochastic}),
but they showed the power of the technique through several examples.  The paper
has since been cited over six thousand times.

Conditions for ergodicity of Gibbs samplers and Metropolis-Hastings algorithms are given in \cite{tierney1994markov}
and simpler, less-general conditions for convergence of the Gibbs sampler are given in \cite{roberts1994}.
Gibbs sampling and many other MCMC algorithms are described in the text book by
Jun Liu \cite{liu2008monte}.  This includes blocked Gibbs samplers (also referred
to as ``grouped") and other variations. The textbook ``Bayesian Data Analysis" (BDA) \cite{gelman2014bayesian} contains
examples of Bayesian regression models and Gibbs samplers tailored to them.
BDA also includes examples of blocked Gibbs samplers (e.g. chapter 15 section 5 in the third edition).

Blocked Gibbs sampling was evaluated in the context of Gaussian Mixed Effects models
by Chib and Carlin \cite{chib1999mcmc}.  They considered the Gaussian longitudinal model:
\begin{equation}
y_k = X_k\beta + W_kb_k + \epsilon_k \in \mathbb{R}^{n_k}
\end{equation}
where $b_k \in \mathbb{R}^q$ and has Gaussian prior $b_k \sim N_q(0, \Sigma)$. 
They developed seven different Gibbs samplers with varying levels of blocking. 
These included samplers which integrated out $\{b_k\}$ and $\beta$ and drew
$\Sigma$ conditional on only the observed data $\{ y_k \}$.  They also showed
how to apply their algorithms to binary probit regression using the latent
variable representation $z_k = \text{sign}(y_k)$ where $z_k$ is observed but
$y_k$ is not.  They found that blocking substantially reduced autocorrelation
in their examples that the additional computational cost of the blocked updates
was a good tradeoff.

In our experience the blocked updates are especially important when there is a
long tail of levels which have little associated data.  Consider advertisers as an example:
there may be a small subset of advertisers responsible for a huge number of ``views" and ``actions",
and these are most informative when learning the prior variance parameters $\sigma$.
Often though, there is also a much larger population of advertisers with very sparse data,
and their presence slows down the mixing of a non-blocked Gibbs sampler.

The work of Volfosky and Hoff \cite{volfovsky2012hierarchical} is similar in that
they build models with many random effect families: their focus is on Gaussian ANOVA models
with multiple factors and interaction terms.  They implement Gibbs samplers
of balanced designs and suggest data augmentation as a technique to handle
imbalanced designs (i.e. add missing data which would make the design balanced).
They focused on relatively small datasets (i.e. $n < 10,000$, $F < 3$, and $L_k < 10$).
The algorithm we describe below has been applied to much larger problems.

In the next section we give a detailed description of the algorithm and associated computations, and
in section \ref{sec:exts} we discuss some extensions to the algorithm.

\section{Gibbs Sampling for Gamma-Poisson Regression Models \label{sec:GammaPoisson}}
In this section we give a detailed description of the blocked Gibbs sampler for
the Gamma-Poisson model.  In an un-blocked Gibbs
sampler with variables $(V_1, V_2, ..., V_n)$ we would iterate the updates:
\begin{equation}
v_k \sim p(V_k|V_1=v_1, ..., V_{k-1}=v_{k-1}, V_{k+1}=v_{k+1}, ..., V_n = v_n)
\end{equation}

In a ``blocked" or ``grouped" Gibbs sampler we can update several variables at a
time:
\begin{equation}
\{v_j:j\in I\} \sim p(\{V_j:j\in I\}|\{V_i:i\not\in I\} = \{v_i:i\not\in I\})
\end{equation}

We describe how the block $I$ can be taken to be a large set of random effects
as well as an associated prior variance parameter.  This blocked update is just
as scalable as unblocked updates (and much more scalable than a naive implementation
of an unblocked Gibbs sampler).

As mentioned earlier, we do not develop an algorithm for general $X$ and $Z$
model matrices.  We will handle the case that $X$ is the $n\times 1$ matrix $[1]_{n\times 1}$ and so $\beta \in \mathbb{R}$.
With the restriction on $Z$ described above we can more compactly write the
subsequent computations in terms of an $n\times F$ matrix of indices $I$.  We define $I_{jk} = t$ if
$Z_{j, (t+T_{k-1})} = 1$.  For example, if the first family of random effects is
based on ``url" and we enumerate the unique url's as 1,2,3,..., then $I_{j1} = t$ if
the $j$'th row of the input table contains the $t$'th url.  To make some
equations easier to read we will sometimes write $I(j,k)$ in place of $I_{jk}$.

The table below shows what the matrix $I$ would look like for the dataset in
the introduction (Table \ref{tab:exampledata1}).  The columns with headings $I_1$ and $I_2$ show the first and
second columns of the matrix $I$.   
\begin{center}
  \begin{tabular}{ | l | c | c | c | c | r | }
    \hline
    n.views & n.actions & url & ad.id & $I_1$ & $I_2$ \\ \hline \hline
    52 & 4 & abc.com & 83473 & 1 & 1 \\ \hline
    73 & 5 & xyz.edu & 40983 & 2 & 2 \\ \hline
    19 & 0 & abc.com & 4658 & 1 & 3 \\ \hline
    532 & 16 & efg.com & 40983 & 4 & 2 \\ \hline
    3 & 0 & z.com & 4658 & 5 & 3 \\ \hline
    ... & ... & ... & ... & ... & ... \\ \hline
    \hline
  \end{tabular}
\end{center}

Next we will define $\RgdB_{kt} := \exp(b_{T_{k-1} + t})$.  $B$ is a ragged array
rather than a matrix because the length
$\{\RgdB_{kt}\}_t \in \mathbb{R}^{L_k}$ can depend on the family index $k$.
As a shorthand we will simply write $\RgdB_k := \{\RgdB_{kt}\}_t$ for the vector of random
effects associated with the $k$'th family.

This representation is not just notationally convenient -- it is also how we represent
$Z$ and $b$ in our optimized implementation of the algorithm. Next we define three
operations in terms of componentwise vector products/divisions, and these
operations represent the bulk of the computation in large datasets.  These operations are
used to compute the sufficient statistics which appear in our Gibbs sampling
steps.
\begin{align}
  \Predict(\RgdB, \beta) &:= \beta \Offset \prod_{f=1}^F \BExp{f}{} \in \mathbb{R}^n \label{eqn:PoisPred} \\
  \Predict_{-k}(\RgdB, \beta) &:= \Predict(\RgdB, \beta) / \BExp{k}{} \in \mathbb{R}^n \\
  \SumBy_k(V) &:= \{ \sum_{j:I(j,k)=i}V_j \}_{i=1}^{L_k} \in \mathbb{R}^{L_k}, V \in \mathbb{R}^n  \label{eqn:SumByK}
\end{align}
where $\BExp{k}{} := \{ \RgdB_{k, I(j,k)} \}_{j=1}^n \in \mathbb{R}^n$.  If we
partitioned the columns of $Z$ by family $Z = [Z^{(1)} Z^{(2)} \cdots Z^{(F)}]$ then another
way to define $\BExp{k}{}$ would simply be the matrix-vector product $Z^{(k)}B_k$.  In (\ref{eqn:SumByK})
we use a negative subscript ``$-k$" to remind the reader that the
prediction is based on all but the $k$'th random effect family.

Our data distribution can be rewritten concisely in terms of $\Predict()$:
\begin{align}
  \{Y|\Offset,\RgdB,\beta,Z\} \sim \text{Poisson}(\Predict(\RgdB, \beta))
\end{align}
recalling that $I$ is just a different representation of the matrix $Z$.

To update $\RgdB_k$ we will perform two computations:
\begin{align}
  \events &:= \SumBy_k(Y) \in \mathbb{N}^{L_k} \\
  \pevents &:= \SumBy_k(\Predict_{-k}(\RgdB, \beta)) \in \mathbb{R}^{L_k}
\end{align}

``$\SumBy$" can be computed in $O(n)$ time, and we will show that although the cost of ``$\Predict$" is $O(nF)$, this can be reduced through amortization.

Our blocked Gibbs sampling update for the $k$'th family is then:
\begin{align}
  \sigma_k &\sim P(\sigma_k|Y,\Offset,\RgdB_{-k}, \sigma_{-k}, \beta) && \text{(integrating out $\RgdB_k$)} \label{eqn:integprior} \\
  \RgdB_k &\sim P(\RgdB_k|Y,\Offset,\RgdB_{-k}, \sigma, \beta) && \text{(conditioning on $\sigma_k$)}
\end{align}
This two stage process produces a sample from the joint posterior:
\begin{align*}
  (\RgdB_k, \sigma_k) &\sim P(\RgdB_k, \sigma_k|Y,\Offset,\RgdB_{-k}, \sigma_{-k}, \beta)
\end{align*}
Due to conditional independence this update is independent of $\sigma_{-k}$
\begin{align*}
  (\RgdB_k, \sigma_k) &\sim P(\RgdB_k, \sigma_k|Y,\Offset,\RgdB_{-k}, \beta)
\end{align*}

It is important to note that the conditional distributions will only depend on $2L_k$ sufficient statistics:
 $\events \in \mathbb{N}^{L_k}$ and $\pevents \in \mathbb{R}_{>0}^{L_k}$.
This is also true for the Gaussian regression described in section \ref{sec:GaussReg}.
\footnote{This simplification does not occur for logistic models
($P(Y_k = 1) = (1 + \exp(-\theta))^{-1}$) or truncated Gaussian regressions which
are used in a latent-variable representation of binary probit regressions \cite{chib1999mcmc}.}

A second point worth noting is that we need not precisely sample the variance parameter
in equation (\ref{eqn:integprior}); to maintain the correct stationary distribution 
it is sufficient to use Metropolis-Hastings \cite{van2014metropolis}.

Earlier we mentioned that it was possible to speed up the computation of the
$\Predict$ functions defined above, and we describe this now.  Suppose that
$\ModelPred^{old} := \Predict(\RgdB, \beta)$ was computed before sampling
$(\RgdB_k, \sigma_k)$.  Once we have drawn a new value $\RgdB_k^{\new}$ we can update:
\begin{align*}
  \ModelPred^{\new} &:= \ModelPred^{old} \frac{\BExp{k}{\new}}{\BExp{k}{old}} \in \mathbb{R}^n && \text{(componentwise multiplication/division)}
\end{align*}
The un-amortized cost of the $\Predict_{-k}(B, \beta)$ operations
would be $O(nF^2)$ for a single scan over all $F$ families.  In practice we compute
$\Predict(\RgdB, \beta)$ using equation (\ref{eqn:PoisPred}) at the beginning of
each scan to reduce the cost by a factor of $F$.  Due to accumulation of
numerical errors we cannot refresh once per scan for arbitrarily large $F$; however,
we have observed no practical consequences when applying this to models with $F$
in the hundreds.

If we let $\ModelPred := \Predict_{-k}(B, \beta)$ then data log likelihood for rows $j$ with
index $I_{jk}=t$ is:
\begin{align*}
  \ell_t(\RgdB_{kt}) &:= \sum_{j: I(j,k)=t} (-\RgdB_{kt}\ModelPred_j + Y_j\log(\RgdB_{kt}\ModelPred_j) - \log(Y_j!)) \\
    &= -\pevents_t \RgdB_{kt} + \events_t \log(\RgdB_{kt}) + c(t, Y, \ModelPred)
\end{align*}
where $c(t, Y, \ModelPred) := \sum_{j: I(j,k)=t} (Y_j\log \ModelPred_j - \log(Y_j!))$

The prior likelihood on $B_{kt}$ is
\begin{align*}
  P(\RgdB_{kt}=u|\sigma_k) &= \dgamma(u, \sigma_k^{-2}, \sigma_k^{-2}) \\
\text{where } \dgamma(x, \theta, \eta) &:= C_\Gamma(\theta, \eta) x^{\theta - 1}\exp(-\eta x) \\
\text{and } C_\Gamma(\theta, \eta) &:= \frac{\eta^{\theta}}{\Gamma(\theta)}
\end{align*}

Because the prior on $\RgdB_{kt}$ is a product of independent Gamma distributions,
each element of $(\RgdB_{kt}, \{ Y_j: I(j,k)=t \})_t$ is independent
after conditioning on $(\Offset, \RgdB_{-k}, \sigma_k, \beta)$.  This
conditional independence is used to simplify some high dimensional integrals
into products of one-dimensional integrals in the formulas below.

Using this conditional independence
we compute the marginal data likelihood after integrating out $\RgdB_k$:
\begin{align}
P(Y|\Offset, B_{-k}, \sigma_k) 
  &= \int \prod_t P(B_{kt}=u_t|\sigma_k)\exp(\ell_t(u_t)) du_1 ... du_{L_k} \label{eqn:margdatadistn} \\
  &= \prod_t \int P(B_{kt}=u|\sigma_k)\exp(\ell_t(u)) du \\
  &= \prod_t \exp(c(t, Y, \ModelPred))\frac{C_\Gamma(\sigma_k^{-2}, \sigma_k^{-2})}{C_\Gamma(\sigma_k^{-2}+\events_t, \sigma_k^{-2}+\pevents_t)}
\end{align}
In the equation above we did not condition on the other prior parameters, $\sigma_{-k}$, or integrate them out because $Y$ is
independent of $\sigma_{-k}$ after conditioning on $B_{-k}$.  For use in subsequent pseudocode we define the function $\PriorMarginal()$ as
\begin{align}
\PriorMarginal(\sigma_k, &\events, \pevents) \nonumber \\
  &:= \prod_t \frac{C_\Gamma(\sigma_k^{-2}, \sigma_k^{-2})}{C_\Gamma(\sigma_k^{-2}+\events_t, \sigma_k^{-2}+\pevents_t)} \label{eqn:priormarg}
\end{align}

To sample $\sigma_k$ we take the product of the marginal data likelihood and the prior on $\sigma_k$
\begin{align*}
P(Y, \sigma_k|&\Offset, B_{-k}) = P(Y|\Offset, B_{-k}, \sigma_k)P(\sigma_k)
\end{align*}
which is proportional to the posterior $P(\sigma_k|Y, \Offset, B_{-k})$ (as a
function of $\sigma_k$).  We assume a flat, improper prior $P(\sigma_k) \equiv 1$ because we have
no preferred choice.

We do not bother to compute $\prod_t \exp(c(t, Y, \ModelPred))$ since we must
renormalize or use Metropolis-Hastings anyway.  The result is that we need only
compute the aggregate statistics `events' and `pevents' to find
$P(\sigma_k|Y,\Offset, B_{-k})$.

Once $\sigma_k$ is drawn, the posterior distribution of $B_k$ is simply
a product of Gamma distributions:
\begin{align*}
P(B_k|\sigma_k, Y, \Offset, B_{-k})
  = \prod_{t=1}^{L_k}\dgamma(B_{kt}, (\events_t + \sigma_k^{-2}),
                                   (\pevents_t + \sigma_k^{-2}))
\end{align*}

Finally, we discuss the update for the fixed effect parameter $\beta$.  The fixed effect $\beta$ can be
updated through a Monte Carlo EM algorithm, but for simplicity in the pseudocode
we just put a $\text{Gamma}(1, 1)$ prior on $\beta$ and update
it like the other random effects.  We expect little difference in behavior when applied to large
datasets.

The pseudocode in algorithm (\ref{algo:gammapoisson}) summarizes the blocked 
Gibbs sampling algorithm described above.  As mentioned earlier, step \ref{alg:priorupd}
in the algorithm could be a single Metropolis-Hastings update.  The
``griddy Gibbs sampler" updates described in \cite{tierney1994markov} are another option.
Finally, in subsequent pseudocode we write just ``Sample $B_{f}^{\new} \sim P(B_f|B_{-f}, Y, \Offset, \beta, \sigma)$"
in place of the for-loop on lines \ref{alg:ranefupd1}-\ref{alg:ranefupd2}.

\begin{algorithm}
\caption{Gamma-Poisson Gibbs Sampling algorithm}\label{algo:gammapoisson}
\begin{algorithmic}[1]
\State initialize $B$, $\beta$, $\sigma$
\For{iter = $1,2,...$}
  \State $\ModelPred \leftarrow \Predict(B, \beta)$
  \State Sample $\beta^{\new} \sim \dgamma(\beta, 1 + \sum_j Y_j, 1 + \beta^{-1}\sum_j \ModelPred_j)$
  \State $\ModelPred \leftarrow \ModelPred \beta^{\new} / \beta$
  \State $\beta \leftarrow \beta^{\new}$  \Comment{Update $\Pi$ for use in the next sampling step}
  \For{$f = 1,2,..., F$}\Comment{For each feature family}
      \State $\events \leftarrow \SumBy_f(Y)$
      \State $\pevents \leftarrow \SumBy_f(\ModelPred / \BExp{f}{})$
      \State Sample $\sigma_f$ from $\PriorMarginal(\sigma_f, \events, \pevents)$ \label{alg:priorupd}
      
      \For{$t = 1,2,..., L_f$}\Comment{Sample $B_{f}^{\new} \sim P(B_f|B_{-f}, Y, \Offset, \beta, \sigma)$} \label{alg:ranefupd1}
         \State $B_{ft}^{\new} \sim \dgamma(\events_t + \sigma_f^{-2}, \pevents_t + \sigma_f^{-2})$
      \EndFor  \label{alg:ranefupd2}
      \State $\ModelPred \leftarrow \ModelPred \frac{\BExp{f}{\new}}{\BExp{f}{}}$  \Comment{Update $\ModelPred$ for use in the next sampling step}
      \State $B_f \leftarrow B_f^{\new}$
  \EndFor
\EndFor
\end{algorithmic}
\end{algorithm}

\section{Extensions \label{sec:exts}}
In this section we discuss extensions which remain as scalable as the
Gamma-Poisson regression described in the previous section.

\subsection{Handling A More General $Z$ Matrix \label{sec:GeneralZ}}
Suppose the matrix $Z$ from the introduction can be partitioned into two sets of
columns $Z = [Z^{(1)} | Z^{(2)}]$ and that $Z^{(1)}$ is structured as we required for algorithm
\ref{algo:gammapoisson}.  If on the other hand $Z^{(2)}$ is not structured this way,
then clearly we can use the algorithm \ref{algo:gammapoisson} to update the
random effects and priors associated with $Z^{(1)}$ and use more general
updates for $Z^{(2)}$.

We take a moment to discuss one seemingly straightforward extension that turns
out to not be as scalable in the Poisson model.  Suppose the elements of $Z$ were not 0-1 but still had
the sparsity pattern $\sum_{t \in J_k}1\{Z_{it} \neq 0 \} = 1$.  We can represent
the information in $Z$ compactly using two $n \times F$ matrices $I$ and $S$.  While the
index map $I$ stores the sparsity pattern, the additional matrix $S$ stores the non-zero values of $Z$  i.e. $S_{jk}$ is equal
to the $(j, T_{k-1} + I_{jk})$'th element of $Z$ (recall the definition of $T_k$
 in section \ref{sec:intro} was $\sum_{j=1}^k L_k$).

In this more general model we would modify $\Predict$ to instead be:
\begin{equation}
  \ScaledPredict(B, \beta) := \beta \Offset \prod_{f=1}^F \BExp{f}{}^{S_f} \in \mathbb{R}^n \label{eqn:scalepred}
\end{equation}
where $S_f = \{S_{jf} \}_{j=1}^n$ is the $f$'th column of $S$ and the exponentiation $\BExp{f}{}^{S_f}$ is
taken componentwise.

Under this generalization the vectors `{\events}' and `{\pevents}' are no longer the sufficient statistics
for the conditional distribution of $(B_k, \sigma_k)$.  We must instead aggregate
by unique values of $(S_{jk}, I_{jk})$ rather than of $I_{jk}$.

Unless $S_f$ takes on few unique values, the vector of sufficient statistics can be as long as $2n$
elements.  Another important difference is that the conditional distributions of
the elements of $B_f$ will no longer have a Gamma distribution.

\subsection{More Flexible Prior Distributions}
In order to efficiently implement our blocked Gibbs sampler it is important
that we can compute the marginal data likelihood
$\int P(B_{kt}=u|\sigma_k)\exp(\ell_t(u)) du$ in closed form.  This is
possible because the Gamma distribution is a conjugate prior for the
Poisson distribution.

We still have considerable flexibility because these integrals can be
evaluated analytically for mixtures of discrete and Gamma distributions as well:
\begin{align*}
  P(B_{kt}& \in A) = \sum_{j=1}^d w^k_j 1\{\text{loc}^k_{j} \in A\}
      + \sum_{j=d+1}^{d+g} w^k_j \int_A \dgamma(x, \sigma_{kj}^{-2}, \sigma_{kj}^{-2})dx \\
\text{where }& \text{loc}^k_{j} \in \mathbb{R}_{\geq 0}, \sigma_{kj} \in \mathbb{R}_{> 0}, \sum_j w^k_j = 1 
\end{align*}

A common special case would be a sparse prior:
\begin{align*}
  P(B_{kt} \in A) &= w^k 1\{1.0 \in A\} + (1 - w^k)\int_A \dgamma(x, \sigma_{k}^{-2}, \sigma_{k}^{-2})dx
\end{align*}
This is referred to as a ``spike and slab" or ``spike and bell" prior
\cite{george1993variable}, \cite{mitchell1988bayesian}.

Sampling from the two-dimensional conditional distribution of $(w^k, \lambda_k)$
may require more care to implement, but the likelihood is still a function of $2L_k$
statistics.

\subsection{Gaussian Regression Models \label{sec:GaussReg}}
When defining the Gaussian model, we will refer to the matrix $S$ defined in
equation (\ref{eqn:scalepred}).  In that section we pointed out that handling a more general
$Z$ matrix came at significant computational cost in the Poisson model; however,
in the Gaussian model this is not the case.  The updates are equally simple
after lifting the 0-1 restriction on the entries of $Z$.  Again we consider
the case with sparsity pattern $\sum_{t \in J_k}1\{Z_{it} \neq 0 \} = 1$ for $k=1,2,...,F$.

The {\Predict} functions are similar to those defined for the Poisson model:
\begin{align}
  \GaussPredict(B, \beta) := \beta + \sum_{f=1}^F \BExp{f}{}S_f \in \mathbb{R}^n \label{eqn:gausspred} \\
  \GaussPredict_{-k}(B, \beta) := \beta + \sum_{f \neq k} \BExp{f}{}S_f \in \mathbb{R}^n \label{eqn:gausspred2}
\end{align}
where $\BExp{f}{}S_f$ is a componentwise product.  Our model of the data
is now:
\begin{align}
  Y \sim N(\GaussPredict(B, \beta), \Offset^{-1})
\end{align}
where $\Offset$ now serves as an residual inverse-variance rather than an
offset in the regression.  We do not develop the sampling steps needed to
infer the residual variance in this report.

We next define the sufficient statistics for the Gibbs sampling steps:
\begin{align}
  \invvar &:= \SumBy_k(S_k^2\Offset) \in \mathbb{R}^{L_k} \label{eqn:invvar} \\
  \error &:= \SumBy_k((Y - \GaussPredict_{-k}(B, \beta))S_k\Offset) \in \mathbb{R}^{L_k}
\end{align}
All operations are taken to be componentwise -- including the squared term $S_k^2$ in equation (\ref{eqn:invvar}).

As in the Poisson model we will compute the data log likelihood associated with
each level of the random effect:
\begin{align*}
  \ell_t(B_{kt}) &:= -(1/2)\sum_{j: I(j,k)=t} \left((Y_j - \ModelPred_j - S_{jk}B_{kt})^2\Offset_j + \log(2\pi / \Offset_j) \right) \\
    &= -(1/2)\invvar_t B_{kt}^2 + \error_t B_{kt} + c(t, Y, \ModelPred)
\end{align*}
where $c(t, Y, \ModelPred) := -(1/2)\sum_{j: I(j,k)=t} \left((Y_j - \ModelPred_j)^2\Offset_j + \log(2\pi / \Offset_j) \right)$

We place a $N(0, \sigma_k^2)$ prior on $B_{kt}$, and, as in
equation (\ref{eqn:margdatadistn}), we will integrate over the random effects:
\begin{align*}
P(Y|\Offset, B_{-k}, \sigma_k)
  &= \int \prod_t P(B_{kt}=u_t|\sigma_k)\exp(\ell_t(u_t)) du_1 ... du_{L_k} \\
  &= \prod_t \int P(B_{kt}=u|\sigma_k)\exp(\ell_t(u)) du \\
  &= \prod_t \exp(c(t, Y, \ModelPred))\left(\frac{\sigma_k^{-2}}{\invvar_t + \sigma_k^{-2}} \right)^{1/2}\exp\left( \frac{1}{2}\frac{\error_t^2}{\invvar_t + \sigma_k^{-2}} \right)
\end{align*}
We did not condition on $\sigma_{-k}$ or integrate it out because $Y$ is
independent of $\sigma_{-k}$ after conditioning on $B_{-k}$.  For the purpose of pseudocode we will define:
\begin{align*}
\GaussPriorMarginal(&\sigma_k, \error, \invvar) \nonumber \\
  &:= \prod_t \left(\frac{\sigma_k^{-2}}{\invvar_t + \sigma_k^{-2}} \right)^{1/2}\exp\left(\frac{1}{2}\frac{\error_t^2}{\invvar_t + \sigma_k^{-2}} \right)
\end{align*}

This unnormalized likelihood is sufficient to update $\sigma_k$ and can be computed from
$2L_k$ sufficient statistics.  Once $\sigma_k$ is drawn, the random effects can
be drawn from their Gaussian posterior distributions:
\begin{align*}
P(B_k|Y,\Offset,B_{-k}, \sigma_k) = \prod_{t=1}^{L_k} \dnorm\left(B_{kt}, \frac{\error_t}{\invvar_t + \sigma_k^{-2}}, (\invvar_t + \sigma_k^{-2})^{-1/2} \right)
\end{align*}
where $\dnorm(x, \mu, s)$ is the normal density with mean $\mu$ and variance $s^2$.
pseudocode for the Gaussian model is given in algorithm \ref{algo:gauss}.

Finally we should note that the computational complexity of the updates (in terms
of $n$ and $L_k$) is unchanged if we generalize to say
$S_{jk}, B_{kt} \in \mathbb{R}^2$ and let $\sigma_k^2$ denote
a $2\times 2$ covariance matrix. The update for $\sigma_k$ would depend on a vector of $3L_k$
sufficient statistics rather than a vector of length $2L_k$, and the sampling
step for each $B_{kt}$ would be to draw from a bivariate rather than a univariate
normal distribution.

\begin{algorithm}
\caption{Gaussian Gibbs Sampling algorithm}\label{algo:gauss}
\begin{algorithmic}[1]
\State initialize $B$, $\beta$, $\sigma$
\For{iter = $1,2,...$}
  \State $\ModelPred \leftarrow \GaussPredict(B, \beta)$
  \State Sample $\beta^{\new}$ \Comment{Details omitted}
  \State $\ModelPred \leftarrow \ModelPred + \beta^{\new} - \beta$
  \State $\beta \leftarrow \beta^{\new}$  \Comment{Update $\Pi$ for use in the next sampling step}
  \For{$f = 1,2,..., F$}\Comment{For each feature family}
      \State $\invvar \leftarrow \SumBy_f(S_f^2\Offset)$
      \State $\error \leftarrow \SumBy_f((Y - \ModelPred + \BExp{f}{}S_f)S_f\Offset)$
      \State Sample $\sigma_f$ from $\GaussPriorMarginal(\sigma_f, \error, \invvar)$ \label{alg:priorupd}
      
      \For{$t = 1,2,..., L_f$}\Comment{Sample $B_{f}^{\new} \sim P(B_f|B_{-f}, Y, \Offset, \beta, \sigma)$} \label{alg:ranefupd1}
         \State Sample $B_{ft}^{\new} \sim \dnorm\left(B_{ft}, \frac{\error_t}{\invvar_t + \sigma_f^{-2}}, (\invvar_t + \sigma_f^{-2})^{-1/2} \right)$
      \EndFor  \label{alg:ranefupd2}
      \State $\ModelPred \leftarrow \ModelPred + \left(\BExp{f}{\new} - \BExp{f}{}) \right)S_f$
      \State $B_f \leftarrow B_f^{\new}$
  \EndFor
\EndFor
\end{algorithmic}
\end{algorithm}

\subsection{Monte Carlo Expecation Maximization (MCEM) algorithm}
We briefly describe the EM algorithm and discuss the similarities with the Gibbs
sampling algorithm.  Our approach is similar to that in \cite{casella2001empirical}
except that we integrate out some random effects -- paralleling the prior updates we
carried out in the full Bayesian approach described above.

The vanilla MCEM algorithm would generate samples
$B^1, ..., B^T$ by Gibbs sampling only the random effects while leaving
the prior variance parameters $\sigma$ fixed. The M-step then decomposes into
independent one-dimensional optimizations which can be performed in parallel:
\begin{align*}
\sigma_k := \argmax{s} \sum_{i=1}^T \sum_{t=1}^{L_k} \log\left( \dgamma(B^i_{kt}, s^{-2}, s^{-2}) \right)
\end{align*}

In practice this seems to converge much more slowly than the blocked Gibbs sampler
we developed in section \ref{sec:GammaPoisson}.
The vanilla MCEM algorithm behaves like the unblocked Gibbs sampler which updates $\sigma_k$
conditional on $B_k$ rather than integrating over $B_k$.

At this point we depart from the description in \cite{casella2001empirical} and
change the update to mimic the blocked Gibbs sampler.  We instead suggest updating just a single
$\sigma_k$ at a time and remove $B_k$ from the complete data log likelihood
(i.e. integrate over it).  The update requires a sample of $B_{-k}$ from its
marginal posterior, but sampling the complete random effect vector $B$
(including $B_k$) is an easy way to generate this.  Algorithm
\ref{algo:mcemgammapoisson} lists pseudocode which omits the maximum likelihood
update for $\beta$.

\begin{algorithm}
\caption{Gamma-Poisson MCEM algorithm}\label{algo:mcemgammapoisson}
\begin{algorithmic}[1]
\State initialize $B$, $\beta$, $\sigma$
\For{iter = $1,2,...$}
  \For{$k = 1,2,..., F$}
      \For{$s = 1,2,..., T$}
         \State $\ModelPred \leftarrow \Predict(B, \beta)$
         \For{$f = 1,2,..., F$}
            \State Sample $B_f^{\new} \sim P(B_f|B_{-f}, Y, \Offset, \beta, \sigma_f)$
            \State $\ModelPred \leftarrow \ModelPred \frac{\BExp{f}{\new}}{\BExp{f}{}}$
            \State $B_f \leftarrow B_f^{\new}$
         \EndFor
         \State $\pevents^{(s)} \leftarrow \SumBy_k(\ModelPred / \BExp{k}{})$
      \EndFor
      \State $\events \leftarrow \SumBy_k(Y)$
      \State $\sigma_k \leftarrow \argmax{v} \sum_{s=1}^T \log \PriorMarginal(v, \events, \pevents^{(s)})$
  \EndFor
  \State Update $\beta$ (details omitted)
\EndFor
\end{algorithmic}
\end{algorithm}

Algorithm \ref{algo:minimalmcemgammapoisson} reduces the number of sampling
steps in algorithm \ref{algo:mcemgammapoisson} as a tradeoff between accuracy
and efficiency.  It may be surprising, but
we have found in practice that algorithm \ref{algo:minimalmcemgammapoisson}
works well when both $n$ and $L_k$ are large; however, when the number of levels
$L_k$ is small, we have seen that a single sample iteration is not sufficient.
In those cases there is little additional cost in taking a fully Bayesian approach
and sampling $\sigma_k$ from its posterior as well.

\begin{algorithm}
\caption{Minimal Gamma-Poisson MCEM algorithm}\label{algo:minimalmcemgammapoisson}
\begin{algorithmic}[1]
\State initialize $B$, $\beta$, $\sigma$
\For{iter = $1,2,...$}
  \State $\ModelPred \leftarrow \Predict(B, \beta)$
  \For{$k = 1,2,..., F$}
      \State $\pevents \leftarrow \SumBy_k(\ModelPred / \BExp{k}{})$
      \State $\events \leftarrow \SumBy_k(Y)$
      \State $\sigma_k \leftarrow \argmax{v} \left(\PriorMarginal(v, \events, \pevents)\right)$
      \State Sample $B_k^{\new} \sim P(B_k|B_{-k}, Y, \Offset, \beta, \sigma_k)$
      \State $\ModelPred \leftarrow \ModelPred \frac{\BExp{k}{\new}}{\BExp{k}{}}$
      \State $B_k \leftarrow B_k^{\new}$
  \EndFor
  \State $\beta \leftarrow \beta \frac{\sum_j Y_j}{\sum_j \ModelPred_j}$
\EndFor
\end{algorithmic}
\end{algorithm}

\section{Discussion}
First we should address what we mean by ``scalable".  On a single workstation we
have applied these algorithms on data sets with $n$ around $100$ million,
$F$ in the tens, and $L_k$ less than one million for each family.

The computational complexity of each update is $O(nF + L_k)$ i.e. linear in the number of input records and
linear in the number of levels of the random effect. For an entire sequential
scan it would be $O(nF^2 + r)$, recalling that $r=\sum_k L_k$ the total number of
random effects and $F$ is the number of random effect families.  Through the amortization
described in section \ref{sec:GammaPoisson}
we can effectively reduce the cost by a factor of $F$ to $O(nF^2\text{min}(F, M)^{-1} + r)$
where experimentally we have found that we can take $M > 100$.

The ``Consensus Bayes" framework is another scalable approach to Bayesian hierarchical
modeling \cite{rabinovich2015variational}, \cite{scott2013bayes}.  We believe
that it is possible to develop consensus versions of the models
presented above, but we have not pursued this yet.  However, the computations
described above lend themselves to parallelized implementations even without
the use of Consensus Bayes techniques.


\begin{thebibliography}{9}
	\bibitem{bates2014fitting}
	  Bates, Douglas and M{\"a}chler, Martin and Bolker, Ben and Walker, Steve.
	  \emph{Fitting linear mixed-effects models using lme4}.
	  arXiv preprint arXiv:1406.5823,
	  2014.

	\bibitem{casella2001empirical}
	  Casella, George.
	  \emph{Empirical bayes gibbs sampling}.
	  Biostatistics,
	  2(4).  pg 485--500,
	  2001.
  
	\bibitem{chib1999mcmc}
	  Chib, Siddhartha and Carlin, Bradley P.
	  \emph{On MCMC sampling in hierarchical longitudinal models}.
	  Statistics and Computing,
	  9(1).  pg 17--26,
	  1999.
	\bibitem{gelfand1990sampling}
	  Gelfand, Alan E. and Smith, Adrian F.M.
	  \emph{Sampling-based approaches to calculating marginal densities}.
	  Journal of the American Statistical Association, Taylor \& Francis Group,
	  85(410).  pg 398--409, 
	  1990.

	\bibitem{gelman2014bayesian}
	  Gelman, Andrew and Carlin, John B and Stern, Hal S.
	  \emph{Bayesian data analysis}.
	  Taylor \& Francis,
	  2014.

	\bibitem{geman1984stochastic}
	  Geman, Stuart and Geman, Donald.
	  \emph{Stochastic relaxation, Gibbs distributions, and the Bayesian restoration of images}.
	  Pattern Analysis and Machine Intelligence, IEEE Transactions on,
    6 pg 721--741,
	  1984.

	\bibitem{george1993variable}
	  George, Edward I and McCulloch, Robert E.
	  \emph{Variable selection via Gibbs sampling}.
	  Journal of the American Statistical Association, Taylor \& Francis Group,
    88(423). pg 881--889
	  1993.

	\bibitem{liu2008monte}
	  Liu, Jun S.
	  \emph{Monte Carlo strategies in scientific computing}.
	  Springer Science \& Business Media,
	  2008.

	\bibitem{mitchell1988bayesian}
	  Mitchell, Toby J and Beauchamp, John J.
	  \emph{Bayesian variable selection in linear regression}.
	  Journal of the American Statistical Association, Taylor \& Francis Group,
    83(404). pg 1023--1032
	  1988.

	\bibitem{rabinovich2015variational}
	  Rabinovich, Maxim and Angelino, Elaine and Jordan, Michael I.
	  \emph{Variational consensus Monte Carlo}.
	  arXiv preprint arXiv:1506.03074,
	  2015.

	\bibitem{roberts1994}
	  Roberts, Gareth O. and Smith, Adrian F.M.
	  \emph{Simple conditions for the convergence of the Gibbs sampler and Metropolis-Hastings algorithms}.
	  Stochastic Processes and their Applications,
    49(2). pg 207-–216,
	  1994.

  \bibitem{stan2015rstansoftware}
    \emph{RStan: the R interface to Stan, Version 2.8.0}.
    http://mc-stan.org/rstan.html,
    2015.

	\bibitem{scott2013bayes}
	  Scott, Steven L and Blocker, Alexander W and Bonassi, Fernando V and Chipman, H and George, E and McCulloch, R.
	  \emph{Bayes and big data: The consensus Monte Carlo algorithm}.
	  EFaBBayes 250 conference,
	  16, 
	  2013.

	\bibitem{tierney1994markov}
	  Tierney, Luke.
	  \emph{Markov chains for exploring posterior distributions}.
	  The Annals of Statistics,
	  pg 1701--1728, 
	  1994.

	\bibitem{van2014metropolis}
	  van Dyk, David A and Jiao, Xiyun.
	  \emph{Metropolis-Hastings within Partially Collapsed Gibbs Samplers}.
	  Journal of Computational and Graphical Statistics, Taylor \& Francis,
    {\it accepted},
	  2014.

	\bibitem{volfovsky2012hierarchical}
	  Volfovsky, Alexander and Hoff, Peter D.
	  \emph{Hierarchical array priors for ANOVA decompositions}.
	  Technical Report, University of Washington, Department of Statistics,
	  2012.

	\bibitem{wang2013parallelizing}
	  Wang, Xiangyu and Dunson, David B.
	  \emph{Parallelizing MCMC via Weierstrass sampler}.
	  arXiv preprint arXiv:1312.4605, 
	  2013.

	\bibitem{welling2011langevin}
	  Welling, Max and Teh, Yee W.
	  \emph{Bayesian learning via stochastic gradient Langevin dynamics}.
	  Proceedings of the 28th International Conference on Machine Learning (ICML-11),
	  pg 681--688, 
	  2011.

\end{thebibliography}
\end{document}